\documentclass[submission,copyright,creativecommons]{eptcs}
\providecommand{\event}{FMAS 2021} 
\usepackage{breakurl}             

\usepackage{color,soul}

\usepackage{cite}
\usepackage{graphicx}
\usepackage{textcomp}
\usepackage{xcolor}
\usepackage{hyperref}
\usepackage{amsmath,amssymb,amsfonts,amsthm}
\usepackage{algorithmic}
\usepackage{mathtools}
\usepackage{thmtools}
\usepackage{ae,aecompl}
\usepackage{calc}
\usepackage{framed}
\usepackage{stmaryrd}
\usepackage{bussproofs}
\usepackage{wrapfig}
\usepackage{yhmath}
\usepackage{enumitem}
\usepackage{listings}
\usepackage{tikz}
\usetikzlibrary{arrows,shapes,automata,petri,backgrounds,calc,intersections,decorations.pathreplacing}

\usepackage{styles/roadlogic}
\usepackage{styles/shortcuts}

\definecolor{darkorange}{rgb}{0.7,0.3,0}
\definecolor{darkred}{rgb}{0.7,0.0,0.1}
\definecolor{darkpurple}{HTML}{8A2BE2}
\definecolor{darkblue}{rgb}{0.0,0.1,0.6}
\definecolor{darkgreen}{rgb}{0.0,0.3,0.0}

\title{Extending Urban Multi-Lane Spatial Logic to Formalise Road Junction Rules\thanks{This research was supported by the German Research Council (DFG) in the PIRE Projects SD-SSCPS and ISCE-ACPS under grants no. FR 2715/4-1, FR 2715/5-1.}}

\author{Maike Schwammberger
\institute{University of Oldenburg\\ Oldenburg, Germany}
\email{schwammberger@informatik.uni-oldenburg.de}
\and
Gleifer Vaz Alves
\institute{Federal Univeristy of Technology - Parana\\ Ponta Grossa, Brazil}
\email{\quad gleifer@utfpr.edu.br}
}
\def\titlerunning{Extending UMLSL for Road Junction Rules}
\def\authorrunning{M. Schwammberger \& G. Vaz Alves}
\begin{document}
\maketitle

\begin{abstract}

During the design of autonomous vehicles (AVs), several stages should include a verification process to guarantee that the AV is driving safely on the roads. One of these stages is to assure the AVs abide by the road traffic rules. To include road traffic rules in the design of an AV, a precise and unambiguous formalisation of these rules is needed.
However, only recently this has been pointed out as an issue for the design of AVs and the few works on this only capture the temporal aspects of the rules, leaving behind the spatial aspects. Here, we extend the spatial traffic logic, Urban Multi-lane Spatial Logic, to formalise a subset of the UK road junction rules, where both temporal and spatial aspects of the rules are captured. Our approach has an abstraction level for urban road junctions that could easily promote the formalisation of the whole set of road junction rules and we exemplarily formalise three of the UK road junction rules.
Once we have the whole set formalised, we will model, implement, and formally verify the behaviour of an AV against road traffic rules so that guidelines for the creation of a Digital Highway Code for AVs can be established.
\end{abstract}

\section{Introduction}\label{sec:introduction}  
Even though autonomous vehicles (AVs) are not yet thoroughly used on our roads \cite{jin_tesla_2021}, we are aware that sooner or later we shall see AVs driving on the roads \cite{hawkins_tesla_2021}. We consider autonomous vehicles that comply with SAE levels 4 or 5 \cite{sae2018}, meaning that the vehicle is either completely driverless, or it manages specific manoeuvres driverless, without a human driver intervening at any point.

    So, there is a need to face many challenges. Especially, those issues related to the safety of AVs, i.e. how to assure that the AV behaves safely on the roads?
    For that, several issues, like obstacle avoidance, sensing the environment, speed control, object detection and recognition and the proper use of traffic rules, need to be addressed.
    
    So far, the issue of traffic rules has not been a major concern for the design of an AV in the research community, as discussed by Prakken \cite{prakken_problem_2017} and Alves et al. \cite{alves_formalisation_2020}. 
    However, some recent work like the references \cite{avary_safe_2020}, \cite{philipp_safety_2019} and~\cite{law_commission_automated_2020} 
    have started to draw attention to the challenge of transforming a Highway Code into a Digital Highway Code.
    Notice that a set of traffic rules composes the rule book or precisely the Highway Code, while 
    a Digital Highway Code is the version of the Highway Code supposed to comprehend those traffic rules designed for AVs.
    There is a clear trade-off on how to wrap the traffic road rules into a digital format in a way that the fewest possible changes are made considering the existent Highway Codes \cite{avary_safe_2020}. At the same time, this digital version of the Highway Code should be understandable for the AVs \cite{law_commission_automated_2020}.
    
    However, for a Digital Highway Code that works for AVs, it is necessary to tackle the challenges of translating road traffic rules (written in natural language) into a language understandable for autonomous systems. Such language needs to be precise and 
    unambiguous since these rules are involved in the process of safety assurance of road users. Once these rules are formalised and deployed into an AV, the AV behaviour can be properly checked against road traffic scenarios to assure that safety road requirements are being followed by the AV (NB: here safety requirements are only those related to the road traffic rules).
    
    In this paper, we follow ideas related to previous work (see ref. \cite{jsan10030041}), where the UK Highway Code (specifically the section of Road Junction rules \cite{department_for_transport_using_2017}) has been used as a basis for the proof-of-concept presented in ref. \cite{jsan10030041}.
For the road traffic rules from the UK Highway Code, temporal and spatial aspects can be identified. For instance, ``look all around \textbf{before} entering the junction''; ``do not cross a road \textbf{until} there is a \textbf{safe gap}'' (``before'' and ``until'' reveal a temporal aspect, while ``safe gap'' reveals a spatial aspect).
As a consequence, we need a formalism suitable to abstract not only the temporal aspects but also the spatial elements of the road traffic rules. \emph{Linear temporal logic (LTL)} is a clear answer to capture the temporal aspects, and was used in references \cite{alves_formalisation_2020, jsan10030041}, to represent the temporal elements of the road junction rules. 
Ref. \cite{jsan10030041} presents an architecture for modelling, implementing, and formal verifying the behaviour of an agent (representing an AV) against three road traffic rules (from the UK Highway Code). 

As a proper candidate to represent spatial elements of traffic rules we identify \emph{\UMLSLL{} (\UMLSL)}, which is used to formalise traffic situations at intersections in \cite{Sch18-TCS}. \UMLSL{} is an interval logic that bases on Interval Temporal Logic (ITL) from \cite{Mos85} and is thus dedicated to capture spatial aspects of traffic. Also, \emph{\eta{} (\acta)} are presented as a formal semantics for a crossing controller for turn manoeuvres. We aim to extend the logic \UMLSL{} from \cite{Sch18-TCS} so that the road junction rules from the UK Highway Code can be formalised and analysed with it.

Our key goal is to enrich the approach from \cite{alves_formalisation_2020, jsan10030041} so that not only the temporal order between events, but also spatial aspects, e.g. a safe gap, can be formalised. For this, we introduce formalisations for non-autonomous traffic participants and road side units (e.g. a traffic sign) and dedicated traffic rule controllers to the approach from \cite{Sch18-TCS}.

Our contribution is organised as follows. As a background, we present an overview over the specification of temporal aspects of traffic rules from \cite{jsan10030041} and give an overview over the spatial traffic logic \UMLSL{} from \cite{Sch18-TCS} in Sect.~\ref{sec:preliminaries}. We motivate and define our \UMLSL{} extension for traffic rules in Sect.~\ref{sec:extension} and exemplarily formalise some of the UK traffic rules in Sect.~\ref{sec:tr-controllers}. We present related work to our approach in Sect.~\ref{sec:related-work} and conclude our work in Sect.~\ref{sec:conclusion} with a summary and some insights into future work possibilities.

\section{Background}\label{sec:preliminaries}
We give preliminary information about the approach on formalising traffic rules using temporal logic from \cite{alves_formalisation_2020} in Sect.~\ref{sec:pre:uk-rules} and in Sect.~\ref{sec:pre:umlsl}, we present details on the abstract model and logic \emph{\UMLSLL{} \UMLSL} from \cite{Sch18-TCS}.
\subsection{The Road Junction Rules}\label{sec:pre:uk-rules}

    In the UK Highway Code there are different sections which handle the road traffic rules for
    Overtaking, Roundabouts, Road Junctions, among others \cite{department_for_transport_using_2017}. 
    Here we are concerned with the section of Road Junction rules, which is composed by 14 rules, from rule 170 to 183. 
    The road junction rules describe how the driver is supposed to behave when entering a road junction, turning to right or left, 
    waiting for a traffic light, etc. 
    As it follows we show the first three rules (170, 171, and 172) that we have been previously formalised in LTL \cite{alves_formalisation_2020} and subsequently used in our agent-based architecture \cite{jsan10030041}.
    Observe that LTL can be used for specifying temporal properties and it uses basic propositional operators ($\land$, $\lor$, $\to$, $\neg$) 
    and temporal modalities ($\square$, $\lozenge$, $\bigcirc$, $\cup$, 
		  representing resp. always, eventually, next, and until).

    \textbf{Rule 170} (UK Highway Code): 
    You should 
	      watch out for road users (\texttt{RU}).
            Watch out for pedestrians crossing a road junction (\texttt{JC}) into which you are turning. 
	If they have started to cross they have priority, so give way.
	Look all around before emerging 
		({NB: For the sake of clarity, we choose to 
		use the term \texttt{enter} as an action which represents not only a driver entering a road junction, but 
		also emerging from a road junction to another road}).
                  Do not cross or join a road until there is a safe gap (\texttt{SG}) large enough for you to do so safely.
	
	\textbf{Rule 170}, represented in LTL, describes when the autonomous vehicle 
	(\texttt{AV}) may enter  the junction (\texttt{JC}):

	\begin{center}

    $\square$ \texttt{((watch(AV, JC, RU)}  $\land$  \texttt{(}$\neg$ \texttt{cross(RU, JC)} $\land$  \texttt{(exists(SG, JC)))}

            $\to$ \texttt{((exists(SG, JC)}  $\land$  $\neg$ \texttt{cross(RU, JC))}  $\cup$ \texttt{enter(AV, JC))))}

	\end{center}

        \texttt{Informal Description}: it is always the case that 		
		the \texttt{AV} is supposed to watch for any road users (\texttt{RU}) at the junction (\texttt{JC}) 
		and there are no road users crossing the junction and there is a safe gap (\texttt{SG}). 
		Then, no road users crossing the junction and the existence of a safe gap should remain true, until 
		the \texttt{AV} may enter the junction.

    \textbf{Rule 170} represented in LTL, when the autonomous vehicle (\texttt{AV}) should give way at the junction (\texttt{JC}):

	\begin{center}

      	$\square $ \texttt{(watch(AV,JC,RU)} $\land$ (\texttt{cross(RU,JC))} $\to$ \texttt{give-way(AV,JC))}           

	\end{center}

        \texttt{Informal Description}: it is always necessary 
		to watch out for road users (\texttt{RU})  and  check if there is a road user 
		crossing the junction. Then, the \texttt{AV} should give way to traffic.

    \textbf{Rule 171} (UK Highway Code): You MUST stop behind the line at a junction with a `Stop' sign (\texttt{ST}) and a solid white line across the road. 
    Wait for a safe gap (\texttt{SG}) in the traffic before you \mbox{move off.}

    \textbf{Rule 171} represented in LTL:

	\begin{center}
			\texttt{exists(ST,JC)} $\to$ $\square$ \texttt{(stop(AV,JC)} $\cup$ \texttt{(exists(SG,JC)}

					 $\land$ \texttt{(exists(SG,JC)}  $\cup$  \texttt{enter(AV,JC))))}			  
	\end{center}

        \texttt{Informal Description}: when there is a stop sign (\texttt{ST}), then it is always the case the \texttt{AV}
       should stop at the junction until there is a safe gap (\texttt{SG}). And the safe gap must remain true until the \texttt{AV} 
       enter at the junction.

    \textbf{Rule 172} (UK Highway Code): The approach to a junction may have a `Give Way' sign (\texttt{GW}) or a triangle marked on the road (\texttt{RO}).
    You MUST give way to traffic on the main road (\texttt{MR}) when emerging from a junction with broken white lines (\texttt{BWL})
    across the road.

    \textbf{Rule 172} represented in LTL:

	\begin{center}
	$\square $ \texttt{((exists(AV,RO)}  $\land$  \texttt{enter(AV,JC))} 

	$\land$  \texttt{((exists(BWL,JC)} $\lor$  \texttt{exists(GW,JC))}
	            $\to$  \texttt{give-way(AV,MR)))}
	\end{center}

    \texttt{Informal Description}: It is always the case that when there is an \texttt{AV} driving on a Road (\texttt{RO})
    and the \texttt{AV} enters the junction and there is a Broken White Line (\texttt{BWL}) or a Give Way sign (\texttt{GW}), then 
			  the \texttt{AV} should give way to the traffic on the Main Road (\texttt{MR}).

\subsection{An Abstract Model for Urban Traffic Scenarios}\label{sec:pre:umlsl}
We introduce the \emph{\UMLSLL{} (\UMLSL)} of \cite{Sch18-TCS} which allows for the formalisation of traffic manoeuvres at intersections. The term \emph{intersection} is equal to the term \emph{road junction} that is used in the UK Highway Code and in \cite{alves_formalisation_2020} (cf.Sect.~\ref{sec:pre:uk-rules}). Hitherto, no traffic rules have been considered using \UMLSL. Nonetheless, some road junction rules are already expressible with it ``by accident''. E.g. safety in the sense of collision freedom has been formally proven in \cite{Sch18-TCS, Sch18, BS19} through mathematical proofs and UPPAAL model-checking \cite{BDL04}.

\begin{figure}[htbp]
\centering
	\includegraphics[scale=.75, trim=0 1.65cm 0 0.9cm, clip]{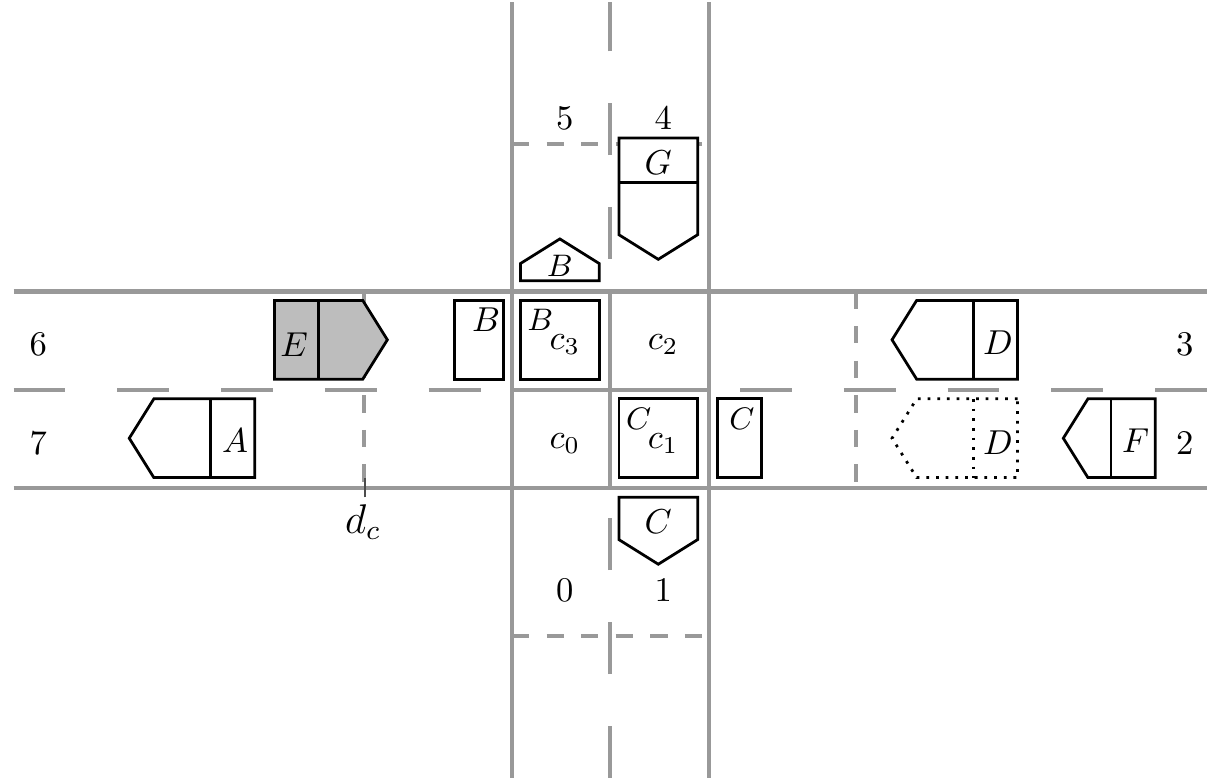}
	\caption{Example for the abstract model from \cite{Sch18-TCS}.}
	\label{fig:model}
\end{figure}

Formulae of \UMLSL{} are evaluated over an abstract representation of real-world intersections. Thus, we first introduce details about this \emph{abstract model} before giving details on the logic \UMLSL{} itself. We focus on those concepts from \cite{Sch18-TCS} that we actually extend in Sect.~\ref{sec:extension} and we leave out formal definitions for most of the concepts in this section. We refer the interested reader to \cite{Sch18-TCS} for more formal and in-depth details for our basis. As a running example, we use the traffic situation that is depicted in Fig.~\ref{fig:model}.

The abstract model contains a set \bigcrossingsegment{} of \emph{crossing segments} $c_0 , c_1 , \ldots$ and a set \Lanes{} of \emph{lane segments} $0, 1, \ldots$ that connect different crossings. Each crossing segment and each lane segment has a finite length. 
Each car is assigned a unique \emph{car identifier} $A, B, \ldots\in\ID$ and a real value for the \emph{position}\index{position} $\pos$ of its rear on a lane or crossing segment. For our example from Fig.~\ref{fig:model}, we use car $E$ as the \emph{\ego{} car} with a valuation $\nu (\ego) = E$ to refer to this car. We distinguish between the \emph{reservation} of a car that formalises the space a car is actually occupying (cf.~$\laneReservation (D)=\{3\}$) and the \emph{claim} of a car, indicating the space a car plans to drive on in the future (cf.~$\laneClaim (D)=\{2\}$, where car $D$ plans to change back to lane $2$ after it finished overtaking the slower car $F$). A claim is thus comparable to setting the turn signal. We also differentiate between claims and reservations on lane segments ($clm$, $res$) and on crossing segments ($cclm$, $cres$).

\textbf{Urban road network.}\label{note:road-network}
Connections of lane and crossing segments are formalised by a directed graph structure called \emph{\roadnetwork{}} \net{} with the set of nodes $\Nu = \Lanes \cup \bigcrossingsegment$. The directed edges between lane and crossing segments specify the driving direction for continuous lane segments. For instance, while a car is allowed to drive from lane $6$ onto crossing segment $c_3$, this is not allowed the other way around.
Each car $C\in\ID$ follows an infinite path $\pth (C)$ with $\pth : \ID\rightarrow (\Z \rightarrow \Nu)$, resembling its travelling route through the \roadnetwork. E.g. in Fig.~\ref{fig:model}, the path of car $E$ for turning right at the depicted crossing is given by $\pth(E) = \langle \ldots 6, c_3, c_2, c_1, 1, \ldots \rangle$.

\textbf{Traffic snapshot.}
Information like the road network \net, reservations, claims, positions and paths of all cars are collected in a global \emph{\traffics{}} \Road. For the example from Fig.~\ref{fig:model}, we have $clm(E)=cclm(E)=\emptyset$, as no space on a lane or crossing segment is claimed for car $E$ (only car $D$ has an active claim $clm(D)=\{2\}$). Further on, we observe $cres(E)=\emptyset$ and $res(E)=\{6\}$ as car $E$ does not occupy a crossing segment but has some space reserved on lane $6$. Car $B$, currently turning at the intersection, has reserved lanes $res(B)=\{5,6\}$ and a crossing reservation $cres(B)=\{c_3\}$.

One \traffics{} can be compared to one snapshot of the overall traffic at an intersection at one moment. Whenever, e.g., time passes or a car claims or reserves a new lane or crossing segment, the \traffics{} changes with respective \traffics{} evolution transitions.
For instance, with a time transition $\Road_0 \xrightarrow[]{t} \Road_1$ a \traffics{} $\Road_0$ evolves to a \traffics{} $\Road_1$, meaning that new positions are determined for all cars $C\in\ID$ after $t$ time units passed and cars moved along their paths with respect to their speed and acceleration values.
Other \traffics{} evolution transitions are triggered by the cars themselves. E.g., with a transition $\Road_0 \transition{\claimCrossing{E}} \Road_1$, crossing segments are claimed for car $E$ along its path through the intersection.

\textbf{Virtual view.}\label{note:virtual-lane}
For reasoning about traffic manoeuvres with the two-dimensional logic \emph{\UMLSLL{} (\UMLSL)}, it is unrealistic and moreover unnecessary to consider an arbitrarily large \traffics{} \Road. Instead, we consider only a finite excerpt of \Road{} called \emph{Virtual View} (cf.~\cite{WW07}). A virtual view $V(E)=(L,X,E)$ is built around the \ego{} car $E$ and contains a sequence of parallel \emph{virtual lanes} $L$ and an extension interval $X$ that determines how far ``ahead'' and ``back'' car $E$ looks. For the example from Fig.~\ref{fig:model} and for a right-turn view $V(E)$ for car $E$, we have virtual lanes $L=\langle \langle 6,c_3, c_2, c_1,1 \rangle , \langle 7, c_0 ,0\rangle \rangle$.

\textbf{\UMLSLL.} Formulae of \UMLSL{} are built from (spatial) atoms, Boolean connectors and first-order quantifiers. Further on, spatial concepts that are inspired by \emph{Interval Temporal Logic} (ITL) \cite{Mos85} are used.
\UMLSL{} introduces four different types of spatial atoms; The atom $re(C)$ (resp.~$cl(C)$) formalises the reservation (resp. claim) of an arbitrary car $C$ on some lane or crossing segment. With the atom \free, free space on a lane or crossing segment is formalised and $cs$ represents crossing segments. Note that no differentiation between between a crossing claim or reservation and a lane claim or reservation is done on the syntactical level of \UMLSL. Also note that the lane number of a reserved lane is not available on the syntactical level of the atom $re(C)$. This is as the goal of this atom is neither to specify the identifier of a reserved lane nor the exact position of a car $C$ on that lane, but rather to formalise whether a lane exists on which car $C$ has a reservation.
By combining these atoms with Boolean connectors, we can, e.g., state that car $E$ occupies a crossing segment ($cs \wedge re(E)$) or that a crossing segment is free ($cs \wedge \free$).

With the spatial connector \chop, \UMLSL{} uses a variation of the chop operator $;$ from ITL. With a spatial formula $re (E) \chop \mathit{free}$, we can, e.g., state that there is free space in front of the reservation of our ego car $E$. Note that ``in front of'' or ``right of'' are informal descriptions for the adjacency of the two space intervals that are formalised by the atoms $re (E)$ and $\mathit{free}$.

Beside the horizontal chop operator \chop, \UMLSL{} also introduces a vertical chop operator which is used by arranging two \UMLSL{} formulae $\phi_1$ and $\phi_2$ one above the other. With this, elements that are located on two neighbouring lane segments can be formalised. E.g., the formula ${}_{cl(D)}^{re(D)}$ describes the situation where car $D$ has a reservation on lane $3$ and a claim on the neighbouring lane $2$.

\UMLSL{} introduces a comparison $u=v$ of variables $u,v \in\variables$ to, e.g., compare two car identifiers and a comparison $\ell = r$, to reason about the length $\ell$ of a spatial interval. This is, e.g., used for checking the distance of a car to an upcoming intersection.
\begin{definition}[Syntax of \UMLSL]\label{def:syntax-umlsl}
Consider a car variable $c \in \carvariables$, a real variable $r \in \realvariables$ and general variables $u,v \in \variables$. The syntax of atomic \emph{\UMLSL{} formulae} is defined by
  \(
    \mathtt{a} ::=  cs \mid \true \mid u=v \mid \ell=r \mid \free \mid \reserved{c} \mid \claimed{c} \text{,}
  \)
whereas an arbitrary \emph{\UMLSL} formula ${\phi}_U$ is formalised as follows:
\abovedisplayskip3pt
\belowdisplayskip3pt
  \begin{align*}
    {\phi}_U ::= \mathtt{a} \mid \lnot \phi \mid \phi_1 \land \phi_2 \mid \exists c \colon \phi_1 \mid \phi_1 \chop \phi_2 \mid {}_{\phi_1}^{\phi_2}
  \end{align*}
   We denote the set of all \UMLSL{} formulae by \({\formulae}_{\mathbb{U}}\).
\end{definition}
In the following, we frequently use the abbreviation $\langle \phi \rangle$\label{note:somewhere} to state that an arbitrary formula $\phi\in{\formulae}_{\mathbb{U}}$ holds \emph{somewhere} in a view $V(E)$ of car $E$. This modality is used to abstract from exact positions in \UMLSL{} formulae.

\begin{example}[Syntax of \UMLSL]\label{ex:syntax-umlsl}
    For the example from Fig.~\ref{fig:model}, the \UMLSL{} formula
    \begin{align}\label{formula:ca}
        ca (E) \,\equiv\, \langle re(E) \chop (\mathit{free} \land \ell<d_c \land \neg cs) \chop cs \rangle
    \end{align}
    formalises the ``crossing ahead check'' for car $E$, meaning that in front of (in Fig.~\ref{fig:model}: ``right of'') the reservation $re(E)$ of car $E$ there is some free space, that is not on an intersection, with a length smaller than $d_c$ and in front of (in Fig.~\ref{fig:model}: ``right of'') that there is a crossing space. 
\end{example}
The logic \UMLSL{} is given a semantics that defines when a traffic snapshot satisfies a given formula. For this, the semantics of a \UMLSL{} formula is evaluated over a \traffics{} \Road, a virtual view $V(E)$ and a variable valuation $\nu$. The variable valuation $\nu$ respects types of variables, so that $\nu \colon \carvariables \rightarrow\ID$ and $\nu \colon \realvariables \rightarrow\mathbb{R}$. Giving formal definitions for the semantics of the basic logic \UMLSL{} would go beyond the scope of formalising traffic rules. However, we explain the semantics of our extension in Sect.~\ref{sec:extension}.

\section{\UMLSL{} for Traffic Rules (\TRUMLSL)}\label{sec:extension}
To formalise traffic rules (cf.~\cite{department_for_transport_using_2017}), we need to be able to reason about traffic signs and non-autonomous traffic participants (e.g. pedestrians, cyclists, human-driven cars, $\ldots$), which is not yet possible using \UMLSL.

Our goal is to keep the necessary \UMLSL{} extension as minimal and elegant as possible. At the same time, we aim for a versatile \UMLSL{} extension that is not tailored around the three traffic rules from Sect.~\ref{sec:pre:uk-rules} that we exemplarily formalise in the following Sect.~\ref{sec:tr-controllers} with our extension. This is as we want to keep the extension as general as possible so that a wider variety of traffic rules is formalisable. This includes that our extension is not limited to UK traffic.

To avoid a cumbersomly long abbreviation like ``UMLSL-TR'', we name the extended logic by \emph{\TRUMLSLL{} (\TRUMLSL)}. \TRUMLSL{} contains all elements of \UMLSL{} and extends its abstract model and logic by two elements:
        \begin{itemize}
            \item A formalism for \textbf{static objects} (i.e. pedestrians, road-side units like traffic signs, traffic lights, $\ldots$), and
            \item a formalism for non-autonomous \textbf{road users} (e.g. cyclists, human-driven cars...).
        \end{itemize}
While it may seem unusual to capture pedestrians within the term ``static objects'', this is a reasonable design decision for the scope of this paper as we explain in the following; One of the main features of the basic logic \UMLSL{} is that formulae of \UMLSL{} are evaluated over a cut-out of an abstract model, which again is built upon a directed graph topology called \roadnetwork. This \roadnetwork{} contains lane and crossing segments that are connected via (un-) directed edges and does not contain sidewalks or a roadside in general in its current version.\label{note:roadside} A semantical introduction of such aspects is non-trivial but seems interesting for future work (cf.~Sect.~\ref{sec:conclusion}). Due to this, we cannot formalise the movement of a pedestrian, e.g. that a pedestrian on a sidewalk ``is about to cross a road''. Thus, for now, we capture a pedestrian that is about to cross a road at a crosswalk or already does so with an abstract static object which is either present or not in one \traffics{} \Road. Informally, this can be compared to a virtual cross-walk that appears whenever a pedestrian wishes to cross a road and that reserves the whole width of the road for crossing pedestrians. 
With this, we can, e.g., formalise the fragment ``Watch out for pedestrians crossing a road junction'' from Rule 170 of the UK Highway Code (cf.~Sect.\ref{sec:pre:uk-rules}). Note that this paper's goal is to formalise traffic rules and that we do not to reason about collision avoidance strategies with pedestrians. For the latter, e.g., movement directions of pedestrians would need to be considered in future work.

Also note that the described design decision implies that we deviate from the term \emph{road user} that was used in the UK traffic rule book \cite{department_for_transport_using_2017} and in the approach that we enrich \cite{alves_formalisation_2020}. Thus, from now on, the term road user comprises non-autonomous and autonomous entities that are not only crossing a road, but are actually able to drive on lane and crossing segments, e.g. cyclists, (non-) autonomous cars, motorcyclists, $\ldots$. We frequently abbreviate the term ``autonomous road user'' to AV. 
In the following, we first describe two approaches that inspire our work in Sect.~\ref{sec:extension:rel-work}. After that, we define the necessary extensions to the abstract model for urban traffic in Sect.\ref{sec:extension:ts} and then introduce syntax and semantics for the new logic \TRUMLSL{} in Sect.~\ref{sec:extension:umlsl}.

\subsection{From Hazards to Road-Side Units and Road Users}\label{sec:extension:rel-work}
The extension \TRUMLSL{} is inspired by two previous approaches that are presented in \cite{OS17, Bi18}. Both approaches introduce moving or stationary hazards to \UMLSL 's predecessor logic \MLSL{} from \cite{HLOR11} to allow for hazard warning protocols. \MLSL{} focuses solely on highway traffic, i.e. one-way traffic and no road intersections. Thus, our contribution is to adapt the ideas from \cite{OS17, Bi18} to the urban traffic case. The term ``stationary hazards'' from \cite{OS17} comprises, e.g., a road accident, dense fog or a damaged road and the term \emph{``moving hazards''} is used in \cite{Bi18} for non-autonomous, human-driven, cars. The main goal of both works is to show that a car receives a hazard warning message early enough and that no collisions with a hazard occur.
Basically, we broaden the term ``hazard'' to a larger variety of objects to formalise traffic rules. We describe key differences and adaptation ideas in the following for both works \cite{OS17, Bi18}.

In \cite{OS17}, the authors introduce an object, namely a single \emph{stationary hazard}, to the highway logic \MLSL. The key difference from our approach is that \cite{OS17} is tailored to cope with multi-lane highway scenarios and not with urban intersections. The second difference is that only one single hazard is allowed for the entire world and that this single hazard is hard-coded into the \traffics{} \Road. To formalise traffic rules, we allow for an arbitrary number of road-side units in one \traffics{} \Road{} and we add a possibility to add new and delete outdated static objects to a \traffics{} \Road. E.g., the need to install a warning sign for a damaged road might exist after an accident occurred but becomes obsolete after the damage was repaired.

In \cite{Bi18}, the author proposes adaptations of \cite{OS17} that allow for multiple stationary and moving hazards on a highway. We adopt a function from \cite{Bi18} that allows for an AV to turn into a moving hazard and vice versa. This is motivated by reality, as it allows for a take-over by a driver, e.g. if the AV has a malfunction that blocks an autonomy function or simply because only some types of manoeuvres can be handled autonomously by the AV (cf.~SAE level 4).

Besides the domain ``urban traffic'', a key difference of our approach from \cite{Bi18} is that we do not give stationary objects a positive extension on a lane or crossing segment. This is because we assume road-side units to be positioned beside the road, not on it and because collisions with road-side units are not a topic of this paper. Further on, we are not restricted to ``hazards'', but instead consider a larger variety of static objects and non-autonomous road users.

\subsection{Changes to the Abstract Model for Urban Traffic}\label{sec:extension:ts}
\begin{figure}[htbp]
\centering
	\includegraphics[scale=.75, trim=0 0.6cm 0 2.15cm, clip]{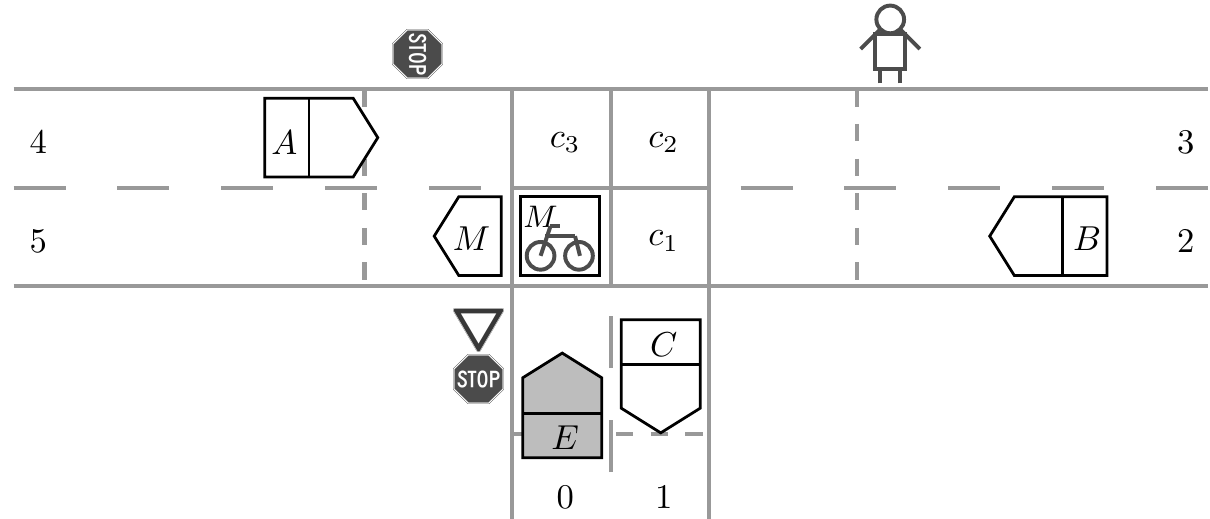}
	\caption{Example for an abstract model with both static objects (stop and give-way sign, a pedestrian), a non-autonomous road user (cyclist $M$) and autonomous road users $A$, $B$ and $E$.}
	\label{fig:model2}
\end{figure}

We explain how to integrate static objects and road users into the existing abstract model for urban traffic from \cite{Sch18-TCS}. Throughout the remainder of this paper, new concepts are explained using the traffic situation that is depicted in Fig.~\ref{fig:model2}.

We introduce the set \OB{} containing identifiers for \emph{static objects} like roadside units into our model. E.g., a stop sign could be identified with $\mathit{stop}\in\OB$. To include static objects $o\in\OB$ into the \traffics{} \Road, we introduce a function \obj, which assigns a set of tuples containing a position and a lane or a crossing segment to each object $o$. With this, the same type of object can exist more than only once in the overall \traffics{} \Road{} (cf. the two priority signs in Fig.~\ref{fig:model2}).
Note that, as motivated in the previous section, we do not assign a positive extension (``size'') to static objects. This means that in our case an object is a dot with a position on the road. This also holds for objects that would have a positive extension in reality, like the pedestrian depicted in Fig.~\ref{fig:model2}. The intuition is that some of the traffic rules, e.g. rule 170 from the UK Highway Code (cf.~Sect.~\ref{sec:pre:uk-rules}), demand that a car should ``watch out for pedestrians crossing a road junction'' and that for this, it is sufficient that an autonomous car realises that there exists a pedestrian at a position at the roadside. However, for future work it might be of interest to add objects with an extension to our model (cf.~Sect.~\ref{sec:conclusion}).

We include identifiers for road users into the existing set \ID of car identifiers and name \ID{} by \emph{set of identifiers} in the following. With this we follow the intuition of \cite{Bi18}, which is to allow for an AV $C$ to turn into a non-autonomous road user and vice versa.

Note that, in our abstract model, we do not formally distinguish between different types of non-autonomous road users and that the visualisation of a bicycle for road user $M$ is only depicted in Fig.~\ref{fig:model2} as a visual reminder that not all road users are AVs as before in Sect.~\ref{sec:pre:umlsl} in Fig.~\ref{fig:model}. This differentiation is not necessary as those three UK traffic rules that were introduced in Sect.~\ref{sec:pre:uk-rules} and that we formalise in Sect.~\ref{sec:tr-controllers} also do not differentiate between different road users. E.g., rule 170 says to ``watch out for road users'' in general. However, for future work it is of interest to distinguish between different road users. For instance, a cyclist may move with a slower velocity than a motorcyclist.

We do not repeat the lengthy definition of \traffics{} elements that was introduced for \UMLSL{} in \cite{Sch18-TCS}. Instead, we only define our object and road user extensions to the \traffics{} \Road{} and abbreviate other \traffics{} elements with $\star$. Such other elements include, e.g., (crossing) reservations, the \roadnetwork{} \net, and positions of cars (cf.~Sect.~\ref{sec:pre:umlsl}).

\begin{definition}[Traffic Snapshot Extensions]\label{def:ts-new}
We extend the Definition of a \traffics{} from \cite{Sch18-TCS} and abbreviate the extension with $\Road = (\star, obj, aut)$, where $\star$ summarises those elements of \Road{} which are defined in \cite{Sch18-TCS} and which are not altered by this definition. Given an arbitrary road user identifier $C\in \ID$ and a static object $O\in\OB$ the new elements in \Road{} are defined as follows:
\begin{itemize}
  \item $\obj \colon \OB \rightarrow \mathcal{P}((\Lanes\cup\bigcrossingsegment) \times \R)$ such that $\obj(O)$ yields a set of 2-tuples of each a lane resp. crossing segment $s\in\Lanes\cup\bigcrossingsegment$ together with a real position of $O$ on the respective segment $s$ and
  \item $\mathit{aut} \colon \ID \rightarrow \mathbb{B}$ indicates whether an element $C\in\ID$ is an AV or a non-autonomous road user.
\end{itemize}
\end{definition}
\begin{example}[Extended Traffic Snapshot]\label{ex:ts-new}
Let us consider the traffic situation that is visualised in Fig.~\ref{fig:model2} and let us assume that the give way sign $\mathit{gw}\in\OB$ is placed close to the intersection on lane $0$ at position $98$ (exemplarily assuming that lane $0$ is, e.g., $100$ units long and that thus $98$ is indeed ``close to the intersection''). We then have $\obj (\mathit{gw}) = \{(0,98)\}$. For the stop signs, which are placed at lanes $0$ and $4$ respectively, we set $\obj (\mathit{stop}) = \{(0,98),(4,198)\}$, exemplarily assuming that lane $4$ is longer than lane $0$ and that the stop sign at lane $0$ is installed at the same position as the give way sign. Note that no positions for stop signs on the neighbouring lanes $1$ and $5$ are provided, as these are both lanes leaving away from the intersection. For the pedestrian $ped\in\OB$, we set $\obj (\mathit{ped}) = \{(2,90),(3,90)\}$ as we assume that for crossing the road, both lane segments $2$ and $3$ are reserved for her as a virtual cross-walk.

For the road users, we have $\aut(M)=0$ for the cyclist $M$ and $\aut(A)=\aut(B)=\aut(E)=1$ for the autonomous cars $A$, $B$ and $E$. The reserved spaces of all road users $A$, $B$, $E$ and $M$ are assigned according to the definition of a \traffics{} \Road{} from \cite{Sch18-TCS}: We have lane reservations $\laneReservation (A) = \{4\}$, $\laneReservation (B) = \{2\}$, $\laneReservation (E) = \{0\}$, $\laneReservation (M) =\{5\}$ and a crossing reservation $\crossingReservation (M) = \{c_0\}$ for road user $M$ on crossing segment $c_0$.
\end{example}

With Def.~\ref{def:ts-new}, we define the extended structure of one single \traffics{} with objects and road users at one distinct moment. As introduced before, a \traffics{} changes, e.g. when a car $C\in\ID$ reserves some crossing segments or when time passes and new positions for all road users are determined.

For static objects $O\in\OB$, we introduce a function $\mathsf{place}$ assigning a new tuple containing a lane or crossing segments and a position to $O$. Reversely, a previously placed object $O$ can be removed from the \traffics{} \Road{} through a function $\mathsf{rm}$. Note that we use the overriding notation $\oplus$ of the specification language \textsf{Z} for function updates \cite{WD96}.
\begin{definition}[Placing and removing static objects]\label{def:ts-evolution1}
\allowdisplaybreaks
Consider a current \traffics{} $\Road = (\star, \obj, \aut)$, where $\star$ again marks those \traffics{} elements that were introduced in \cite{Sch18-TCS} and that are not of concern for this definition. For all $O\in\OB$, $s\in\Lanes\cup\bigcrossingsegment$ and $p\in\R$ the following transitions hold:
\begin{align*}
  \Road \transition{\place{O}{s}{p}} & \Road^{\prime} &\Leftrightarrow
                    &&\Road^\prime &= (\star, \obj^{\prime},\aut) \,\,\,\wedge\,\,\, \obj^{\prime} = \obj \,\,\,\oplus\,\,\, \{O \mapsto \obj (O) \cup {(s,p)}\} \} \\
  \Road \transition{\remove{O}{s}{p}} & \Road^{\prime} &\Leftrightarrow
                    &&\Road^\prime &= (\star, \obj^{\prime},\aut) \,\,\,\wedge\,\,\, \obj^{\prime} = \obj \,\,\,\oplus\,\,\, \{O \mapsto \obj (O) \backslash \{(s,p)\} \} \\
\end{align*}
\end{definition}
\vspace{-0.3cm}
\begin{example}[Placing and removing static objects]\label{ex:ts-evolution1}
Again consider the example from Fig.~\ref{fig:model2}. Through a function call $\place{\stop}{2}{50}$, the visualised \traffics{} evolves as a new instance of the stop sign \stop{} is placed at lane segment $2$ at position $50$. Note that the set union operator ensures that existing placements of $\stop\in\OB$ are not altered. Alternatively, with a function call $\remove{\stop}{0}{98}$, the instance of the stop sign at lane $0$ at position $98$ is removed, where the set difference operator ensures that only the one instance of \stop{} is removed from $\obj(\stop)$.
\end{example}

For non-autonomous road users $ru\in\RU$, we follow \cite{Bi18} and introduce a switching function that can be used to switch an autonomous car $C\in\ID$ to a non-autonomous road user and vice versa.
\begin{definition}[Switching the status of road users]\label{def:ts-switch}
\allowdisplaybreaks
Consider a current \traffics{} $\Road = (\star, obj, aut)$. For all $C\in\ID$ the following transition holds.
\begin{align*}
  \Road \transition{\switch{C}} & \Road^{\prime} &\Leftrightarrow
                    &&\Road^\prime &= (\star, \obj^{\prime},\aut) \,\,\,\wedge\,\,\, 
                    \aut^{\prime} = \aut \,\,\,\oplus\,\,\, \{ C \mapsto \neg \aut (C) \}
\end{align*}
\end{definition}
\begin{example}[Switching the status of road users]\label{ex:ts-switch}
On calling $\switch{A}$, the status $\aut (A)= 1$ of the AV $A$ i changed to $\aut (A)= 0$. With this, $A$ is considered a non-autonomous road user.
\end{example}

\subsection{Syntax and Semantics of \TRUMLSL}\label{sec:extension:umlsl}
We extend the syntax and semantics of \UMLSL{} by atoms for static objects and non-autonomous road users. For this sets of variables \carvariables{} and \objectvariables{} ranging over identifiers from the set \ID{} and over the set of static objects \OB{} are used. Def.~\ref{def:syntax-umlsl} on the syntax of \UMLSL{} is extended as follows for the syntax of the traffic rule logic \TRUMLSL.
\begin{definition}[Syntax of new \TRUMLSL{} concepts]\label{def:USL-TR}
For an object variable $o \in \objectvariables$ and a car variable $c \in \carvariables$, we extend the atomic \emph{\UMLSL{} formulae} $\mathtt{a}$ from Def.~\ref{def:syntax-umlsl} as follows:
  \(
    \mathtt{a}^{\prime} ::=  \mathtt{a} \mid \objectstat{o} \mid \roaduser{c} \text{,}
  \)
All definitions of binary and spatial connectors, as well as first-order logic quantifiers, to build \TRUMLSL{} formulae $\phi_T$ remain as of Def.~\ref{def:syntax-umlsl}. We denote the set of all \TRUMLSL{} formulae by \({\formulae}_{\mathbb{T}}\).
\end{definition}
\begin{example}[Syntax of new \TRUMLSL{} concepts]\label{ex:USL-TR}
Consider car $A$ and the stop sign that it approaches on lane $4$ in the traffic situation that is depicted in Fig.~\ref{fig:model2}. This situation can be formalised by \(\phi_1 \equiv \reserved{A}\chop\free\chop \objectstat{Stop}\), which informally reads as ``there exists a reservation for car $A$, a part of free space in front of $A$ and after that there exists a stop sign''. The existence of the road user $M$ on the intersection in front of car $E$ can be formalised with
\(\phi_2 \,\equiv\, \langle\reserved{E} \,\chop\, \free \,\chop\, (\roaduser{M} \land \mathit{cs})\rangle\)
, which reads as ``there exists a reservation for car $E$, a part of free space in front of $E$ and after that there exists a road user $M$ that is on a crossing segment''.
\end{example}

For the semantics of a \TRUMLSL{} formula $\phi_T$, a variable valuation $\nu \colon \objectvariables \rightarrow\mathbb{O}$ is used for objects and a valuation $\nu \colon \carvariables \rightarrow\ID$ is used for autonomous and non-autonomous road users. For variables $c,d\in\carvariables$ the semantic difference of the atoms \reserved{c} for a reservation of an autonomous car $\nu(c)$ and $\roaduser{d}$ for the reservation of a road user $\nu(d)$ is the autonomy flag $\aut(c)$ (resp. $\aut(d)$). Thus, together with the semantics of the new atoms for static objects \objectstat{o} and road users \roaduser{c}, we give the changed semantics of the atom \reserved{c}. Note that the autonomy flag is the only change of \reserved{c} compared to \cite{Sch18-TCS} and that we explain the mathematical concepts of the definition in detail subsequently. We again use the \textsf{Z} specification language.

\begin{definition}[Semantics of new \TRUMLSL{} concepts]\label{def:semantics-TRUMLSL}
With respect to a traffic snapshot $\Road$, a virtual view $V = (L,X,E)$ and a valuation of variables $\val$, with $c\in\carvariables$ and $o\in\objectvariables$, the \emph{satisfaction} of the spatial \TRUMLSL{} atoms \reserved{c}, \roaduser{c} and \objectstat{o} is defined as follows:
\allowdisplaybreaks
\begin{align}
\Road,V,\val \,\models\, \reserved{c} \,\,\, \Leftrightarrow\,\,\, 
    &\# L =1 \text{ and } |X| > 0 \text{ and } \, \aut (c) = \mathit{true} \text{ and } \forall s_i \colon L (1); \exists X_i \subseteq X \,\bullet\,\nonumber\\
    & s_i \in \crossingReservation (\val(c)) \cup \laneReservation (\val(c))
    \, \text{and } ( s_i ,X_i )\in seg_{V} (\val(c)) \text { and }  X \subseteq \bigcup_{i=1}^{\# L(1)} X_i\label{formula:re}\\
\Road,V,\val \,\models\, \roaduser{c} \,\,\, \Leftrightarrow\,\,\, 
    &\# L =1 \text{ and } |X| > 0 \text{ and } \, \aut (c) = \mathit{false} \text{ and } \forall s_i \colon L (1); \exists X_i \subseteq X \,\bullet\,\nonumber\\
    & s_i \in \crossingReservation (\val(c)) \cup \laneReservation (\val(c))
    \, \text{and } (s_i ,X_i )\in seg_{V} (\val(c)) \text { and }  X \subseteq \bigcup_{i=1}^{\# L(1)} X_i\label{formula:ru}\\
\Road,V,\val \,\models\, \objectstat{o} \,\,\, \Leftrightarrow\,\,\, 
   &\# L =1 \text{ and } \# L(1) = 1 \text{ and } |X| = 0 \text{ and } \exists s \colon L (1); \exists p \colon \mathbb{R} \,\bullet\,\nonumber\\
    & X=[p,p] \text{ and } (s,p) \in \obj(o)\label{formula:ob}
\end{align}
\end{definition}
To satisfy the atomic formulae $\reserved{c}$ and $\roaduser{c}$, the view $V$ has to be occupied completely by the respective element. This holds if $V$ consists of only one virtual lane ($\# L = 1$) and has a positive extension ($|X| > 0$). As a quick reminder for the reader: Basically, one virtual lane complies with one possible path through an intersection (cf. exemplary virtual lanes for the traffic situation from Fig.~\ref{fig:model} on p.~\pageref{note:virtual-lane}). Then it is checked if for all lane or crossing segments $s_i \in\bigcrossingsegment\cup\Lanes$ that are contained in the one virtual lane $L(1)$, a (crossing) reservation for $\nu(c)$ exists ($s_i \in \crossingReservation (\val(c)) \cup \laneReservation (\val(c))$). With the abstract function $seg_V (\nu(c))$, it is checked that the considered extension interval $X_i$ on segment $s_i$ is visible in $V$ and that all segments in $V$ are completely occupied by $\nu(c)$ (cf. last part of formulae \eqref{formula:re} and \eqref{formula:ru}). Definition \eqref{formula:ob} restricts the view $V$ even further: As a static object is only a dot on a segment, the virtual lane $L(1)$ may only contain one single segment $s\in\bigcrossingsegment\cup\Lanes$ and the interval extension is limited to $X=[p,p]$, where $p$ is the position value that is assigned to $s$ in $\obj(o)$. Note that as single \TRUMLSL{} atoms are always required to completely fill a view $V_i$, the larger view $V(E)$ that is considered around one \ego{} car $E$ to reason about traffic rules generally consists of several smaller views $V_i$.
\begin{example}[Evaluation of \TRUMLSL{} formulae over views]\label{ex:semantics-TRUMLSL}
We continue the previous example \ref{ex:USL-TR} and again consider formula \(\phi_2 \equiv \reserved{E}\chop\free\chop (\roaduser{M} \land \mathit{cs})\), which is parted into three parts using the chop operator $\chop$. As each sub-formula $\phi_2^i$ is evaluated over one sub-view $V_i$, the view satisfying $\phi_2$ can be, e.g., parted into these three sub-views.
\end{example}

\section{Formalisation of UK Road Junction Rules using \TRUMLSL}\label{sec:tr-controllers}
With \TRUMLSL{}, we propose a means to formalise spatial aspects of traffic rules, like e.g. that a stop sign is ahead or that a safe gap is large enough for an AV. As motivated earlier, traffic rules also contain temporal aspects, which were the focus of the previous work \cite{jsan10030041} that we briefly outline in Sect.~\ref{sec:pre:uk-rules}. Here, our means to formalise temporal aspects of traffic manoeuvres are extended timed automata controllers that use formulae of \TRUMLSL{} in guards and invariants. With this, we follow \cite{Sch18-TCS}, where \emph{\etas{} (\acta)} were introduced as an extension of the original timed automata from \cite{AD94} to specify and verify a crossing controller for turn manoeuvres at intersections. 

Our overall endeavour is to fully integrate \TRUMLSL{} into the agent-based approach from \cite{jsan10030041}. For this, it is of interest to consider a combination of the agent-based UPPAAL implementation from \cite{jsan10030041} with the UPPAAL implementation that was done in \cite{Sch18, BS19} to verify the system properties of the crossing controller from \cite{Sch18-TCS}. However, this paper focuses on the extension \TRUMLSL. We exemplarily show-case the usability of \TRUMLSL{} by formalising the UK road junction rules 170 to 172 that were introduced in Sect.~\ref{sec:pre:uk-rules}. We sketch connections between the approaches \cite{jsan10030041} and \TRUMLSL{} in the following paragraph but refer to future work for the actual integration of \TRUMLSL{} into \cite{jsan10030041} (cf.~Sect.\ref{sec:conclusion}).

Besides spatial and temporal aspects, the nature of a \emph{rule} generally requests a certain behaviour, i.e. \emph{action}. E.g., the need to stop the car on encountering a stop sign. For this, \cite{jsan10030041} uses abstract actions like \texttt{enter(AV, JC)} or \texttt{stop(AV,JC)} (cf.~Sect.~\ref{sec:pre:uk-rules}). As a counterpart, \acta{} come with certain \emph{controller actions} which allow to, e.g., set a turn signal at an intersection or accelerate/decelerate an AV. In this section, we introduce \TRUMLSL{} guards and invariants for the detection part of a rule (e.g. ``there is a stop sign''). In \cite{jsan10030041}, abstract actions like 
\texttt{watch(AV,JC,RU)} were used for this.
This paper is about steps towards a Digital Highway Code, thus encoding traffic rules for AVs. Nonetheless, the rules also hold for non-autonomous vehicles as they are from the UK Highway Code.

A non-trivial problem that we face in this section, and that was also sketched in \cite{jsan10030041}, is the problem of accurately translating natural language sentences into an accurate machine-readable and -understandable language. E.g., in the part ``watch out for road users'' of rule 170, it is not specified where and when an AV should do this.\label{natural-language-problem} As the rules are from the road junction part of the UK Highway Code, we assume that they should hold \textbf{at intersections} and, if not specified differently, do so \textbf{always}. The natural language translation problem is outlined in detail in \cite{R11}. One of the difficulties is the often imprecise wording and the ambiguous semantics of some natural language phrases. For now, we explain our formalisation choices as detailed as possible. However, automatic means to extract formal specifications of traffic rules from their natural language counterparts are of interest for future work. E.g., in \cite{Getal14}, the authors automatically extract requirements specifications from semi-formal natural language requirements.

\textbf{Safe gap. }\label{note:safe-gap}
We start with the formalisation of a ``safe gap'' as this is a feature frequently required by several of the considered UK traffic rules. Safe gaps on crossing segments are needed if an AV wants to enter an intersection and safe gaps on lane segments are needed when the AV leaves the intersection. For the specification of safe gaps on lane segments for overtaking manoeuvres, we also refer to \cite{HLO13}, where a predecessor logic of \UMLSL{} for two-way country roads is introduced. In our urban traffic case, safe gaps can be formalised for an \ego{} car $E$ by using the basic \UMLSL{} version from \cite{Sch18-TCS}. We assume that the size of a safe gap is relative to the size $\mathit{size}_E$ of the \ego{} car. This size is retrieved in \cite{Sch18-TCS} through a sensor function. A safe gap of free space for \ego{} car $E$ anywhere, i.e. not necessarily on an intersection, can be formalised with the formula
\begin{align}\label{formula:safe-gap-sw}
    \mathit{sg}(E) \,\equiv\, \free \,\land\, \ell >= \mathit{size}_E\textit{.}
\end{align}
Consequently, a safe gap on an intersection (resp. on a lane) can be specified by adding ``$\land\, cs$'' (``$\land\,\neg cs$'' resp.) to formula \eqref{formula:safe-gap-sw}.
However, for road junction rules, we need to specify that the safe gap is free on the intersection in front of $E$ and not on any arbitrary intersection. Thus, formula~\eqref{formula:safe-gap-sw}, needs to be embedded into a formula that is specified from the perspective of and relatively to the \ego{} car $E$. This is done via the formula
\begin{align}\label{formula:safe-gap}
    \mathit{sg}_I (E) \,\equiv\, \langle (re(E) \land \neg cs) \,\chop\, (\mathit{free} \land \neg cs) \,\chop\, (\mathit{sg} (E) \land cs) \rangle\text{,}
\end{align}
which is an adaptation of the crossing ahead check $ca(E)$ from formula~\eqref{formula:ca}, p.~\pageref{formula:ca}. Formula~\eqref{formula:safe-gap} states that the reservation of the \ego{} car $E$ is on a lane segment before an intersection, that there might be some free space between the AV and the intersection (the car will not stand exactly in front of the intersection when it checks for a safe gap) and then there is a free safe gap (cf.~formula~\eqref{formula:safe-gap-sw}) on the intersection.

\textbf{Rule 170. }
As seen in Sect.~\ref{sec:pre:uk-rules}, rule 170 contains the following four parts:
\begin{enumerate}
    \item You should watch out for road users.
    \item Watch out for pedestrians crossing a road junction into which you are turning. If they have started to cross they have priority, so give way.
	\item Look all around before emerging.
    \item Do not cross or join a road until there is a safe gap large enough for you to do so safely.
\end{enumerate}
As we do not identify any dependencies between these parts, we formalise and describe them separately. Each part is required to hold, which could be, e.g., achieved by a conjunction or by modelling four controllers that run in parallel and ensure that each part holds invariantly.

\textit{Part 1.} The first part of rule 170 demands to \textit{``watch out for road users''} (we assume ``always'', ``at intersections'', cf. previous remark on p.~\ref{natural-language-problem}). For this, we refer to a concept from \cite{Sch18-TCS}, where such a feature is already included in the behaviour of the crossing controller: On approaching an intersection, \ego's controller always checks for \emph{potential collisions} using the \UMLSL{} formula 
\begin{align}\label{formula:pc}
	pc(c) \; \equiv \; c \neq \ego \wedge \langle cl(\ego) \wedge (re(c) \vee cl(c)) \rangle \text{.}
\end{align}
For a car $c$ different than \ego, formula~\eqref{formula:pc} checks for path intersections of \ego's own claim (``path through an intersection'') and the claims or reservations of other road users. If a potential collision with any road user $c$ exists, \ego{} withdraws its claim and only enters the intersection later if no potential collisions are detected. Note that in \cite{Sch18-TCS} such road users $c\in\ID$ only include AVs. However, as in this work non-autonomous road users are included into the set of identifiers \ID{} (cf.~Sect.~\ref{sec:extension:ts}, Def.\ref{def:ts-new}), the potential collision check also applies to non-autonomous road users.

An extension of this part of rule 170, namely rule 221, has been formulated in the UK Highway Code and requires to ``watch out for long vehicles which may be turning at a junction ahead'', as those might need the whole intersection for their manoeuvre. Note that this extension is already included in the above description, as a long vehicle would simply demand all crossing segments for its path through the intersection. With this, \ego{} would also watch out for long vehicles with $pc(c)$. However the \roadnetwork{} \net{} (cf.~Sect.~\ref{sec:pre:umlsl}, p.~\pageref{note:road-network}) needs to be built accordingly to allow for long vehicles.

\begin{figure}
    \centering
    \includegraphics{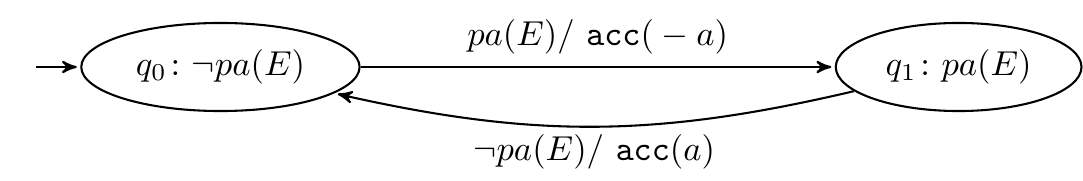}
    \caption{$\acta_{170}$ implementing the part of rule 170 of the UK highway code for road junctions where an AV waits for a pedestrian to cross a road.}
    \label{fig:controller-170}
\end{figure}

\textit{Part 2.} The second part of rule 170 focuses solely on pedestrians: 
\textit{``Watch out for pedestrians crossing a road junction \dots so give way.''}.
To formalise this rule with \TRUMLSL, we again manipulate the crossing-ahead check from formula~\eqref{formula:ca}. Specifically, we formalise a ``pedestrian ahead'' check $pa(E)$ for the \ego{} car $E$:
    \begin{align}\label{formula:pa}
        pa (E) \,\equiv\, \langle re(E) \,\chop\, (\mathit{free} \land \ell< d_p) \,\chop\, \objectstat{Ped} \rangle \text{.}
    \end{align}
Formula~\eqref{formula:pa} holds if a pedestrian $Ped\in\mathbb{O}$ is ahead of the \ego{} car $E$ within a given safety distance $d_p$. Next, the rule demands that the AV should give way to the pedestrian. Basically, \textit{``give way''} is a request that an AV should decelerate or stop to let a pedestrian cross. This is implemented through the simplified traffic rule controller $\acta_{170}$ that we depict in Fig.~\ref{fig:controller-170}: Consider the transition from the initial location $q_0$ to location $q_1$. If the guard~\eqref{formula:pa} holds, \ego{} decelerates using the controller action ``accelerate'' $\mathtt{acc}\,(-a)$ with the negation of an abstract acceleration value $a\in\mathbb{R}^+$ and changes to location $q_1$, where $pa(E)$ holds invariantly. If the pedestrian is not crossing anymore and $\neg pa(E)$ holds, the controller accelerates again with a positive acceleration value $a$. We assume that $a$ is large enough so that the car can decelerate timely and can come to a stillstand before reaching the crossing pedestrian.

Note that we use $\acta_{170}$ as a tool to show how our formalisations can be used as guards and invariants in a controller that is built to follow traffic rules. However, it is a very abstract model for decelerating and accelerating an AV. So, instead of a controller action $acc(a)$ (resp. $acc(-a)$), it is more realistic that $\acta_{170}$ would rather communicate the need for an acceleration (resp. deceleration) to a speed control unit within the AV.

\textit{Part 3.} We interpret that the third part of rule 170, ``Look all around before emerging'', requires the AV to check for path intersections with other AVs, non-autonomous road users or with pedestrians before leaving (``emerging from'') an intersection. This can be formalised by
    \begin{align}\label{formula:look-around}
        \mathit{look} (E) \,\equiv\, \langle (re(E) \land cs) \chop (\neg cs \,\land\, \mathit{sg}(E)) \rangle \,\land\, \neg \exists c \colon \mathit{pc}(c) \land \neg pa(E)\text{.}
    \end{align}
Basically, the three conjugated fragments of formula~\eqref{formula:look-around} formalise more than the informal and imprecise phrase ``Look all around before emerging'' comprises:
\begin{itemize}
    \item it is checked that car $E$ is on an intersection ($re(E) \land cs$) and that after the intersection, there is a free safe gap for car $E$ available for $E$ to emerge into ($\neg cs \,\land\, \mathit{sg}(E)$, cf.~formula~\eqref{formula:safe-gap-sw}),
    \item it is checked that no potential collisions with other road users exist ($\neg \exists c \colon pc(c)$), and
    \item we ensure that no pedestrian is ahead (cf.~formula~\eqref{formula:pa}).
\end{itemize}
If formula~\eqref{formula:look-around} holds as a transition guard in $\acta_{170}$, \ego{} is allowed to leave the intersection, which again means adjusting $E$'s speed. We do not explicitly depict this transition in Fig.~\ref{fig:controller-170}.

\textit{Part 4.} This part demands to ``not cross or join a road until there is a safe gap''. We understand that this rule demands i. that an AV does not enter an intersection (``do not cross'') until there is a fitting safe gap available, and ii. that an AV only then leaves an intersection (``join a road''), if a safe gap is available on the road after an intersection. We already included part ii. into the first fragment of our formalisation~\eqref{formula:look-around}. For part i., we refer back to formula~\eqref{formula:safe-gap}, where we formalised a free safe gap on an intersection that lays ahead of $E$.

\textbf{Rule 171. }
This rule demands that the AV ``must stop behind the line at a junction with a stop sign and a solid white line across the road''. A stop sign or a ``solid white line across the road'' that is ahead can be identified with a \TRUMLSL{} formula equivalent to the pedestrian ahead check from formula~\eqref{formula:pa}. We rewrite the ``pedestrian ahead'' check $pa(E)$ from formula~\eqref{formula:pa} to a ``stop sign ahead'' check $sta(E)$ for the \ego{} car $E$:
    \begin{align}\label{formula:sta}
        sta (E) \,\equiv\, \langle re(E) \,\chop\, (\mathit{free} \land \ell< d_{st}) \,\chop\, \objectstat{\mathit{Stop}} \rangle \text{.}
    \end{align}
Note that we require $\objectstat{\mathit{Stop}}\in\mathbb{O}$ and assume that $d_{st}$ is a distance within which a stop sign is considered close enough for the AV to act. Here, the action again is a deceleration of the AV so that it comes to a timely standstill. Additionally, rule 171 demands to wait for a safe gap before moving onto an intersection (cf.~Sect.~\ref{sec:pre:uk-rules}), which we discussed before in detail for rule 170.

\textbf{Rule 172. }
Instead of a stop sign, a give way sign is in the focus of this rule: ``The approach to a junction may have a `Give Way' sign or a triangle marked on the road. You must give way to traffic on the main road when emerging from a junction with broken white lines across the road''. The identification of a give way sign (or any other optical identification of it) is again similar to the pedestrian ahead check~\eqref{formula:pa}. We name this check by ``give way ahead'', $gwa (E)$, but do not rewrite formula~\eqref{formula:pa} here for reasons of brevity.

If $gwa (E)$ holds, the AV $E$ must give way to the traffic on the main road. For this, we can use a feature from \cite{BS19}, where fairness was introduced to the crossing controller from \cite{Sch18-TCS}. For this, AVs are assigned priorities on arriving at an intersection. The longer an AV waits in front of the intersection, the more its priority increases. For our implementation of rule 172 this means that whenever the ``give way ahead'' formula $gwa(E)$ holds for an AV $E$, $E$ receives a priority penalty. On the other hand, another AV $C$ on the main road receives a priority bonus on approaching a crossing where it has the right of way. With this, using the crossing controller from \cite{BS19}, cars on the main road will always get the right of way.

\section{Related Work}\label{sec:related-work}
There are some approaches which aim to formalise road traffic rules.
Pek et al. \cite{pek_verifying_2017} formalise the safety of lane change manoeuvres to avoid collisions. The authors use as
reference the Vienna Convention on traffic rules to formalise a single rule on the safe distance. 
They use algebraic equation to formalise this road traffic rule.
Rizaldi et al. \cite{Retal17} formalise and codify part of the German Highway Code on the overtaking traffic rules in LTL (three rules are formalised).
They show how the LTL formalisation can be properly used to abstract concepts from the traffic rules and obtain unambiguous and precise specification for the rules.
In addition, they formally verify the traffic rules using Isabelle/HOL theorem prover and also monitor an AV applying a given traffic rule, which has been previously formalised using LTL.
Bhuiyan et al. \cite{bhuiyan_traffic_2020} assess driving behaviour against traffic rules, specifically the overtaking rules from the Queensland Highway Code. Two types of rules are specified: overtaking to the left and to the right.
Moreover, they intend to deal with rules exceptions and conflicts in traffic rules (this is solved by setting priorities among the rules). Using DDL (Defeasible Deontic Logic) they assess the driving behaviour telling if the driver has permission or it is prohibited to apply a given rule for overtaking. The results basically show if the proposed methodology has recommended (or not) the proper behaviour for the driver (permission or prohibition).
Besides,  Esterle et al. \cite{EGK20} present a fomalisation of traffic rules for two-lane roads (``dual carriageways")
in LTL to specify temporal behaviour. A set of formalised traffic rules is presented and evaluated on a public dataset.

In \cite{DMPR18}, Traffic Sequence Charts (TSCs) have been introduced as a visual language for describing first-order logic predicates for traffic situations. TSCs allow to introduce arbitrary objects and traffic rules like, e.g., a lane-change rule are exemplarily formalised in \cite{DMPR18}. However, TSCs abstract from several aspects. Also, we aim at the formalisation of traffic rules at road junctions, while, to our best knowledge, TSCs are limited to multi-lane highway traffic scenarios. Nonetheless, a combination of TSCs with our approach is of interest for future work.
 
The authors of \cite{KD20} focus on the translation of traffic rules from the California's DMV driver handbook from natural language to formal language (i.e. first order logic representations) and simulate their approach for exemplary four way and three way uncontrolled intersections using the CARLA urban driving simulator for autonomous vehicles\cite{CARLA}. They show that the behaviour of autonomous vehicles under their controller are more realistic compared to CARLA's default FIFO controller.
However, neither of the aforementioned works formalise spatial aspects of traffic rules and only one of them \cite{KD20} formalises road junction rules.

\section{Conclusion}\label{sec:conclusion}
We have presented an extension for the logic UMLSL to handle a subset of road junction rules of the UK Highway Code. We capture not only the temporal but also spatial aspects of traffic rules, e.g. a safe gap.

Despite having shown the formalisation of only three road junction rules, the extension of the formalisation for the set of all 14 road junction rules of the UK Highway Code should not pose to many difficulties. This is as the remaining rules outline similar elements such as traffic lights, dual carriageways, other uses of safe gap situations or manoeuvres like turning at a road junction. With static objects and road users, \TRUMLSL{} already contains the necessary concepts to abstract and formalise the aforementioned elements from the road junction rules.

Our vision is that the formalisation can provide some guidelines for the deployment of a Digital Highway Code for AVs. Two important guidelines are i. the spatial abstraction to represent a safe gap, which is largely used throughout the UK road junction rules (cf.~p.~\pageref{note:safe-gap}); and ii. the effortlessly switching of status (autonomous or non-autonomous) for a given road user (cf.~Def.~\ref{def:ts-switch}), which helps to represent special emergency scenarios, where the AV control is given back to the human driver (cf. \cite{alves_reliable_2020}).

As future work, we see two research directions; Firstly, we will examine the need for further extensions of \TRUMLSL. E.g., static objects currently have a position on a segment, but no positive extension (``size''). Such a size could be of interest to properly model pedestrian crosswalks or to integrate hazardous situations into the \TRUMLSL{} world (cf.~Sect.~\ref{sec:extension:rel-work}). Also, it would be of interest to consider an integration of a roadside into the abstract model. With this, e.g., a pedestrian approaching a crosswalk on a sidewalk could be specified. Note that with this, we would also be able to consider the movement of pedestrians on sidewalks and could model pedestrians as a special type of road user instead of the static abstraction that we currently use (cf.~p.~\pageref{note:roadside}).

Secondly, and most importantly, we shall model, implement, and verify the road junction rules following the agent-based architecture defined by Alves et al. \cite{jsan10030041}. For that, we will need to change the model and implementation in a way the spatial elements represented by the USL-TR are properly described. The agent implementation will have to consider an extension in the agent's environment to represent the lanes and crossing segments. For this, the UPPAAL implementation of the abstract model for \UMLSL{} from \cite{BS19,Sch20-thesis} might be of help. Also, the status switching function for an AV can be easily captured by changing the agent's belief (since the agent is implemented using a BDI language, cf. \cite{jsan10030041}).
Besides, we intend to use simulation tools like, e.g., CARLA \cite{CARLA} to evaluate our approach in a setting that is closer to reality.

\bibliographystyle{eptcs}
\bibliography{bib}

\end{document}